\documentclass[12pt]{iopart}
\usepackage{graphicx,amssymb,epsfig}

\begin{document}

\title[Pulses in an inhomogeneously broadened
two-level medium]{Ultrashort pulses in an inhomogeneously broadened
two-level medium: Soliton formation and inelastic collisions}

\author{Denis V. Novitsky}
\address{B. I. Stepanov Institute of Physics, National Academy of
Sciences of Belarus, Nezavisimosti Avenue 68, BY-220072 Minsk,
Belarus} \ead{dvnovitsky@gmail.com}

\begin{abstract}
Using numerical simulations, we study propagation of ultrashort
light pulses in an inhomogeneously broadened two-level medium. There
are two main issues in our study. The first one concerns the
transient process of self-induced transparency soliton formation, in
particularly the compression of the pulse which seems to be more
effective in the case of homogeneous broadening. The second question
deals with the collisions of counter-propagating solitons. It is
shown that the level of inhomogeneous broadening has substantial
effect on elasticity of such collisions.
\end{abstract}


\submitto{\JPB}

\maketitle

\section{Introduction}

Self-induced transparency (SIT) is one of the basic phenomena of
nonlinear optics. This effect resulting in formation of temporal
solitons (so-called $2 \pi$ pulses) in a resonantly absorbing medium
was described in the pioneering papers by McCall and Hahn
\cite{McCall1, McCall2}. There is an extensive literature on this
topic including the classic monograph by Allen and Eberly
\cite{Allen} and a number of reviews (see \cite{Kryukov, Lamb,
Poluektov, Maimistov1990}, to name a few). The continuing study of
SIT is motivated by the fundamental and general character of the
two-level model which is the basic tool for explanation of this
phenomenon. The investigation of the properties of this model based
on the semiclassical Maxwell-Bloch equations is necessary for our
understanding of the nature of light-matter interaction. There is a
number of generalizations, as well, aiming to take into account the
near-dipole-dipole interactions (local-field correction) in the
dense resonant media \cite{Bowd93}, Stark shift of the absorption
line \cite{Afan99}, and few-cycle pulse dynamics in the regime of
invalidity of the rotating-wave approximation \cite{Ziolkowski,
Novit4}. Using the two-level model and its generalizations, the
broad spectrum of SIT studies was performed including invariant
pulse propagation and optical switching in dense media \cite{Bowd91,
Cren92}, quasiadiabatic following analysis \cite{Cren96}, incoherent
solitons \cite{Afan02}, SIT soliton lasers \cite{Kozlov}, SIT
soliton collisions \cite{Shaw, Novit2}, SIT in Bragg reflectors and
photonic crystals \cite{Kozhekin, Kurizki, Novit}, coherent pulse
propagation and SIT effects in doped waveguides and amplifiers
\cite{Nakazawa1, Nakazawa2}, SIT in the presence of Kerr
nonlinearity \cite{Maimistov1983, Kozlov1}, etc.

The aim of this paper is to consider some details of SIT soliton
formation and interaction between solitons in the two-level medium
with inhomogeneous broadening of resonant line. This broadening is
due to the fact that, generally, the frequency of resonant
transition is not the same for all atoms of the medium. The
importance of inhomogeneous broadening can be illustrated by the
example of the so-called area theorem. This theorem governs the
change of the ``area'' $S (z)=(2 \mu/\hbar) \int A(z, t) dt$ of the
pulse propagating in the medium and can be written in the form
\begin{eqnarray}
\tan\frac{S(z)}{2}=\tan\frac{S_0}{2} \exp \left(-\frac{\alpha z}{2}
\right), \label{area}
\end{eqnarray}
where $\mu$ is the component of transition dipole moment parallel to
the polarization vector of the electric field, $A$ the electric
field amplitude, $S_0$ the initial area, $\alpha$ the absorption
coefficient of the medium, $\hbar$ the Planck constant. It is known
that the area theorem (\ref{area}) is strictly valid \textit{only}
for the inhomogeneously broadened media, though the main features of
SIT can be observed in the case of homogeneous broadening as well
\cite{Poluektov}. This was confirmed recently by Yu \textit{et al.}
\cite{Yu} who performed the direct simulations of area change. They
demonstrated that in the case of homogeneous broadening, approach of
the area to the stationary value corresponding to the SIT soliton is
accompanied by slowly damping oscillations. These oscillations
cannot be described by the standard formulae such as (\ref{area}).

The present paper continues previous studies of inhomogeneously
broadened media in comparison to their homogeneously broadened
counterparts. Our attention is directed to the two questions studied
in our recent works \cite{Novit2, Novit3} in the approximation of
homogeneous broadening, viz. the transient process of SIT soliton
formation and the collisions of such solitons. After Sec. \ref{stat}
where the model is described, we discuss these two questions in Sec.
\ref{sol} and \ref{coll}, respectively. The paper closes with the
brief conclusions of the results obtained.

\section{\label{stat}Problem statement}

Light interaction with the two-level medium at every spatial point
is governed by the system of semiclassical Bloch equations for
population difference $W$ and microscopic polarization $R$:
\begin{eqnarray}
\frac{dR}{d\tau}&=& i \Omega W + i R \delta - \gamma_2 R, \label{dPdtau} \\
\frac{dW}{d\tau}&=&2 i (\Omega^* R - R^* \Omega) - \gamma_1 (W-1),
\label{dNdtau}
\end{eqnarray}
where $\Omega=(\mu/\hbar\omega)A$ is the normalized electric field
amplitude (or dimensionless Rabi frequency), and $\gamma_1=(\omega
T_1)^{-1}$ and $\gamma_2=(\omega T_2)^{-1}$ are the rates of
longitudinal and transverse relaxation, respectively. Since we are
interested in consideration of the inhomogeneously broadened medium,
the variables $W$ and $R$ are the functions of the normalized
detuning $\delta=\delta_a+\Delta\omega/\omega$, which is the sum of
$\delta_a=(\omega-\omega_a)/\omega$, the normalized detuning of the
field carrier (central) frequency $\omega$ from the average atomic
resonance, and the term $\Delta\omega/\omega$ which describes the
deviations of the resonance frequency from the average value. The
distribution of two-level atoms over the detunings is governed by
the weight function $g(\delta)$, so that the amplitude of the
macroscopic polarization of the medium is given by
\begin{eqnarray}
p=\mu C \int R(\delta) g(\delta) d\delta, \label{Pmacro}
\end{eqnarray}
where $C$ is the concentration of the two-level atoms.

The one-dimensional Maxwell wave equation for light pulse
propagation in resonantly absorbing medium is
\begin{eqnarray}
\frac{\partial^2 E}{\partial z^2}&-&\frac{1}{c^2} \frac{\partial^2
E}{\partial t^2} = \frac{4 \pi}{c^2} \frac{\partial^2 P}{\partial
t^2}, \label{Max}
\end{eqnarray}
Assuming $E= A \exp [i (\omega t - k z)]$ and $P= p \exp [i (\omega
t - k z)]$, this equation can be represented as the expression for
the dimensionless field amplitude, as follows \cite{Novit2}
\begin{eqnarray}
\frac{\partial^2 \Omega}{\partial \xi^2}&-& \frac{\partial^2
\Omega}{\partial \tau^2}+2 i \frac{\partial \Omega}{\partial \xi}+2
i \frac{\partial \Omega}{\partial
\tau} \nonumber \\
&&=3 \epsilon \left(\frac{\partial^2 \rho}{\partial \tau^2}-2 i
\frac{\partial \rho}{\partial \tau}-\rho\right), \label{Maxdl}
\end{eqnarray}
where $\epsilon=\omega_L/ \omega = 4 \pi \mu^2 C / 3 \hbar \omega$
is the normalized Lorentz frequency which describes the strength of
light-matter coupling, and $\rho$ is the integral in the right-hand
side of (\ref{Pmacro}). In all the equations above, $\tau=\omega t$
and $\xi=kz$ are dimensionless time and distance, respectively;
$k=\omega/c$ the wavenumber; and $c$ the light speed in vacuum. Here
we assume, without loss of generality, that the background
dielectric permittivity of the medium is unity, i.e. we consider the
two-level atoms in vacuum. It is also important to note that in
(\ref{Maxdl}) the slowly varying envelope approximation (SVEA) is
not used.

To solve (\ref{dPdtau})--(\ref{Maxdl}) self-consistently, we apply
the numerical approach which is essentially the same as in our
previous studies \cite{Novit}. At the edges of the calculation
region, we apply the so-called absorbing boundary conditions using
the total field / scattered field (TF/SF) and the perfectly matched
layer (PML) methods \cite{Taflove, Anantha}. One should then
supplement the method, at every time step, with the solution of the
Bloch equations for different detunings with subsequent calculation
of the integral $\rho$ according to (\ref{Pmacro}). Note that,
according to the numerical method, we deal with the evolution of the
total electric field (or \textit{complex} amplitude which contains
all the changes of phase due to propagation), the reflected and
transmitted fields being calculated at the corresponding boundaries
due to the TF/SF procedure. The incident field and the direction of
propagation are initialized at these boundaries as well. In other
words, we do not divide the field into two counterpropagating waves
inside the medium, in contrast to the frequently used approach (see,
for example, \cite{Fleck, Forysiak1, Forysiak2}).

It is known that in the SVEA regime, there are analytical solutions
of coherent pulse propagation obtained with the inverse scattering
theory \cite{Abl1, Kaup, Abl2}. In particular, this method allows to
derive the stationary pulse profile, the temporal dynamics of pulse
area and effect of relaxation. However, the numerical simulations
cover broader spectrum of situations, e.g. they allow to take into
account the local-field correction or deviations from slowly varying
envelope and rotating-wave approximations. The same can be said
about collisions of SIT solitons: as far as we know, there are no
analytical solutions of this problem. Therefore, in this paper, we
use only the numerical approach and leave the analytical attempts
for future considerations. Our choice of one-dimensional
approximation (rather than full 3D simulations) is justified by the
possibility of controlling instabilities and other effects of
transversal beam structure by the manipulations with the aperture of
the optical system \cite{Slusher}.

Let us discuss the main parameters of calculations. To preserve the
generality, all the values are represented in dimensionless form,
the central wavelength of pulse $\lambda=2 \pi c / \omega$ being the
main parameter of normalization. We deal with the ultrashort pulses
of Gaussian shape $\Omega=\Omega_p\exp(-t^2/2t_p^2)$, where $t_p$ is
the pulse duration; such expressions are used as boundary conditions
at the left ($z=0$) and right ($z=L$) boundaries of the medium
according to the TF/SF method. Throughout the paper, the amplitude
of the pulses is measured in the units of the characteristic Rabi
frequency $\Omega_0=\lambda/2 \sqrt{2\pi} c t_p$, which corresponds
to the area equal to $2 \pi$. The pulse duration is measured as a
number $N$ of periods of electric field oscillations $T=\lambda/c$
giving its full width at half maximum (FWHM), namely $t_p=N T/2
\sqrt{\ln 2}$. Since we are interested in the study of coherent
interaction of light with the resonant medium, we assume
$\gamma_1=\gamma_2=0$ in our calculations, i.e. the homogeneous
broadening governed by $\gamma_1$ and $\gamma_2$ is absent.

For the distribution function describing the inhomogeneous
broadening, we use the Gaussian envelope
\begin{eqnarray}
g(\delta)=\frac{1}{\sqrt{\pi} \delta_0} \exp(-\delta^2/\delta_0^2),
\label{gGauss}
\end{eqnarray}
where $\delta_0=\Delta\omega_0/\omega$ is the normalized width of
the distribution, the value $\Delta\omega_0$ is inversely
proportional to the characteristic relaxation time $T_2^\ast$. Such
distributions as that in (\ref{gGauss}) appear, for example, as a
result of Doppler broadening in atomic gases. It is convenient to
express the frequency parameters through the pulse duration as it
was done above for the amplitude $\Omega_0$. Therefore we adopt
$\delta_a=0$ (exact resonance in homogeneous case) and $\omega_L
t_p=0.1$. The latter condition also allows us not to take into
account the so-called local-field correction \cite{Bowd93} which can
be neglected, when $\Omega_0/\epsilon \sim (\omega_L t_p)^{-1} \gg
1$ \cite{Novit1}. Similar normalization can be done for the width of
inhomogeneous broadening which is defined through the value
$\Delta_0=\Delta\omega_0 t_p$, so that the condition $\Delta_0 \ll
1$ (or, equivalently, $t_p \ll T_2^\ast$) corresponds to the case of
the homogeneously broadened two-level medium. The concrete values of
parameters can be obtained if one assumes specific values of the
wavelength $\lambda$ and the number of cycles $N$.

\section{\label{sol}Soliton formation}

\begin{figure}[t!]
\centering \includegraphics[scale=1, clip=]{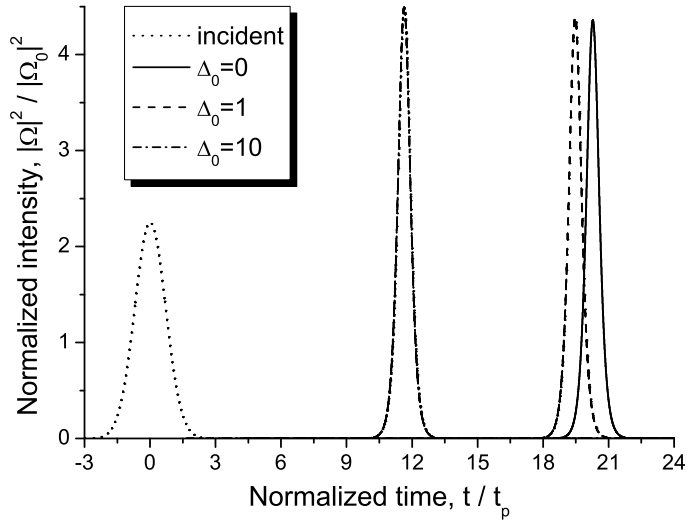}
\caption{\label{fig1} Intensity profiles of the $3 \pi$ pulse
transmitted through the two-level medium with different levels of
inhomogeneous broadening. The thickness of the medium $L=100
\lambda$.}
\end{figure}

First of all, let us consider the influence of inhomogeneous
broadening on the dynamics of self-induced transparency (SIT)
soliton formation. We take very short pulses containing only $N=25$
cycles. Such small duration is convenient from the standpoint of
calculation speed, while, at the same time, it is not too short to
break the rotating-wave approximation used in the Bloch equations.
Anyway, the results can be rescaled for other values of $t_p$. As an
example, let us consider the propagation characteristics of the $3
\pi$ pulse (with amplitude $\Omega_p=1.5 \Omega_0$). Figure
\ref{fig1} shows the results of transmission of such a pulse through
the two-level medium of thickness $L=100 \lambda$. This figure
demonstrates some important aspects to be mentioned here: (i) the
influence of inhomogeneous broadening appears when $\Delta_0 \sim
1$, (ii) increasing $\Delta_0$ leads to the rise in pulse speed in
conformity with the previous studies \cite{Yu}. These facts imply
that our method of calculation works well, so we can directly
proceed to the topic of this paper.

\begin{figure}[t!]
\centering \includegraphics[scale=1, clip=]{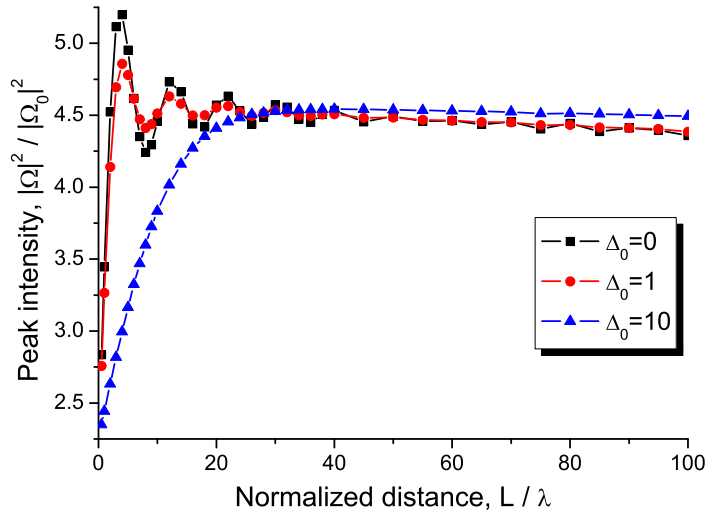}
\caption{\label{fig2} (Colour online) Peak intensity of the $3 \pi$
pulse as a function of distance traversed through the two-level
medium with different levels of inhomogeneous broadening.}
\end{figure}

The dynamics of soliton formation are traced in Fig. \ref{fig2}
where the change in peak intensity of the pulse as it propagates in
the medium is depicted. As previously, the incident pulse has the
area $3 \pi$. The final, quasistationary state (the intensity of the
formed soliton) is almost identical in homogeneous and inhomogeneous
broadening cases, while the initial behaviour differs strongly. The
pulse in homogeneously broadened medium ($\Delta_0=0$) experiences
strong compression soon after incidence and then, through sharp
oscillations of peak intensity, reaches quasistationary, solitonic
form. These oscillations are analogous to those in the area dynamics
studied by Yu \textit{et al.} \cite{Yu} and are characteristic for
this case. Introduction of inhomogeneous broadening (see the curve
for $\Delta_0=1$) makes these oscillations less pronounced, while,
for $\Delta_0 \gg 1$, they are entirely smoothed out. One can
conclude that for smaller broadenings, the pulse can be stronger
compressed during the transient process of soliton formation.

\begin{figure}[t!]
\centering \includegraphics[scale=1, clip=]{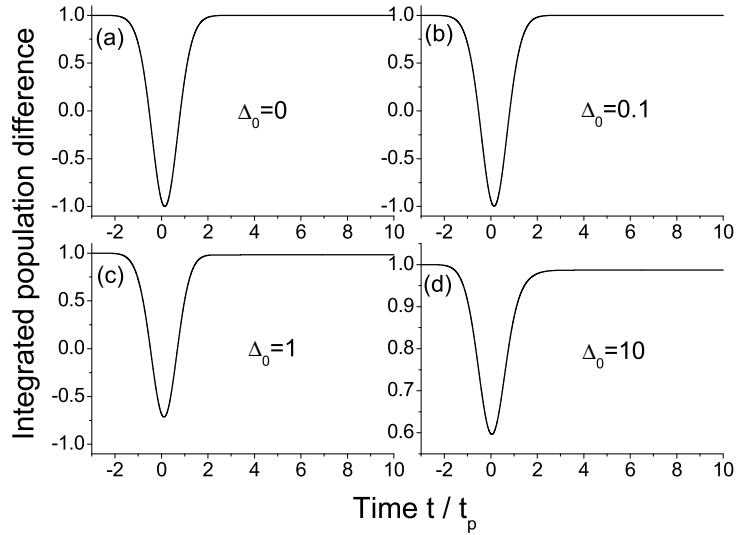}
\caption{\label{fig3} Dynamics of the integrated population
difference $\widetilde{W}$ at the entrance of the two-level medium
with different levels of inhomogeneous broadening. The incident
pulse has the area $2 \pi$.}
\end{figure}

As another measure of inhomogeneous broadening, we will use the mean
value of population difference calculated the same way as $\rho$,
namely
\begin{eqnarray}
\widetilde{W}=\int W(\delta) g(\delta) d\delta. \label{Wmacro}
\end{eqnarray}
In the limit of homogeneous broadening, when all the atoms have the
same resonant frequency, this integrated population difference gives
simply population difference $W$. Under the influence of, say, $2
\pi$ pulse, $\widetilde{W}$ demonstrates typical cycle of inversion
depicted in Fig. \ref{fig3}(a): beginning at the ground state
($\widetilde{W}=1$), the medium switches to the fully inverted state
($\widetilde{W}=-1$) and subsequently returns to the ground one
($\widetilde{W}=1$). The same behaviour is seen in Fig.
\ref{fig3}(b) for $\Delta_0=0.1$ which is effectively still
homogeneous case. As the inhomogeneous width of the spectral line
grows, the energy of the pulse distributes over atoms with different
resonant frequencies, so that the integral value $\widetilde{W}$
cannot reach the full inversion as shown in Fig. \ref{fig3}(c) for
$\Delta_0=1$. For strong inhomogeneous broadening [$\Delta_0 \gg 1$,
see Fig. \ref{fig3}(d)], $\widetilde{W}$ stays near the ground state
at every time instant. In the next section, we will discuss the
implications of these population difference dynamics for collisions
of solitons.

\section{\label{coll}Collisions}

\begin{figure}[t!]
\centering \includegraphics[scale=1, clip=]{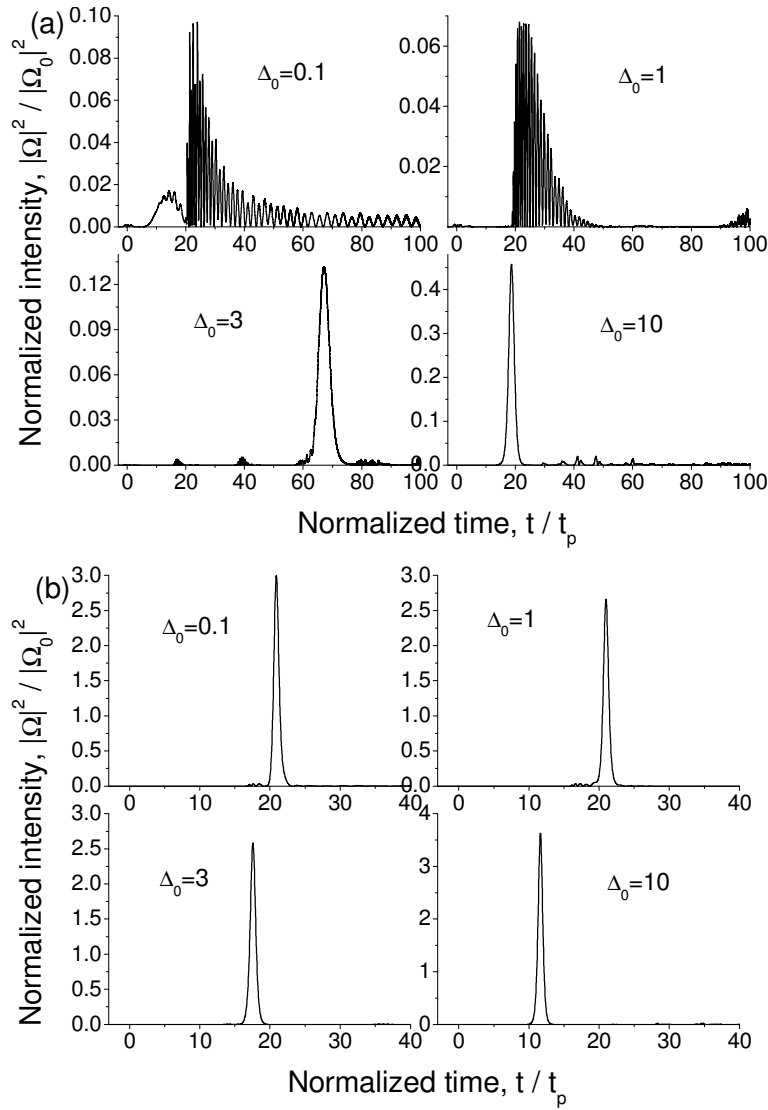}
\caption{\label{fig4} Intensity profiles of (a) the $2 \pi$ forward
propagating (FP) and (b) the $3 \pi$ backward propagating (BP)
pulses after the collision inside the two-level medium with
different levels of inhomogeneous broadening. The thickness of the
medium $L=100 \lambda$. The profiles are recorded at right and left
boundaries of the medium, respectively.}
\end{figure}

In this section, we study influence of inhomogeneous broadening on
the inelastic collisions of counter-propagating pulses in two-level
medium. We are especially interested in the so-called asymmetric
collisions when two colliding solitons are not identical. It was
shown previously \cite{Novit2} that in the case $\Delta_0=0$, such a
collision can result in total destruction of one of the solitons if
the initial parameters of the pulses are chosen properly. Since this
effect is accompanied by strong absorption of light in the point of
collision, it was called the controlled absorption of the soliton
and used as a source of diode action \cite{Novit3}. But what if
$\Delta_0$ is not zero?

\begin{figure}[t!]
\centering \includegraphics[scale=1, clip=]{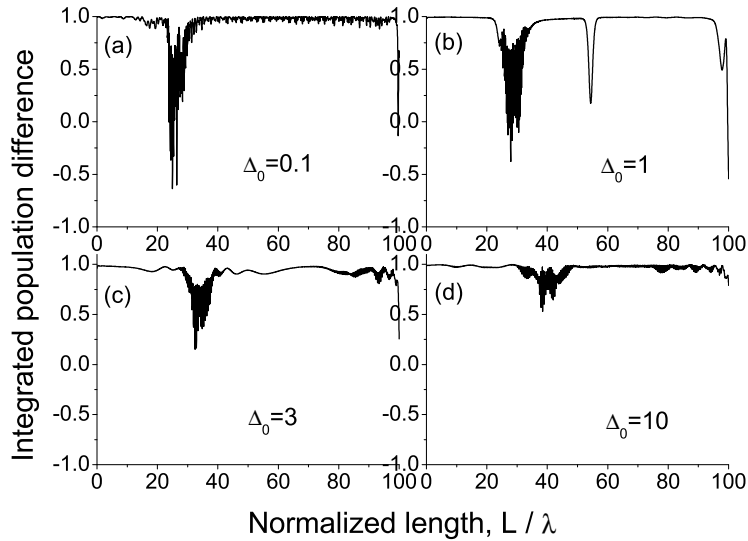}
\caption{\label{fig5} Distribution of the integrated population
difference $\widetilde{W}$ along the two-level medium after the
collisions depicted in Fig. \ref{fig4}. The distributions are
plotted at the instant $100 t_p$.}
\end{figure}

Let us consider the collision of two pulses propagating in the
two-level medium of thickness $L=100 \lambda$: the first,
conventionally called forward propagating (FP, from left to right),
is the $2 \pi$ pulse ($\Omega_p=\Omega_0$), while the second,
backward propagating (BP, from right to left), is the $3 \pi$ pulse
($\Omega_p=1.5\Omega_0$). The results of collisions (profiles of FP
and BP transmitted radiation) for different levels of inhomogeneous
broadening are shown in Fig. \ref{fig4}. One can see that at low
broadenings ($\Delta_0=0.1$ and $1$), the inelastic collision
results in total breakdown of the FP $2 \pi$ pulse: there is no
soliton at the exit of the medium. The radiation contains
low-intensity oscillations, while most part of energy is trapped by
the medium around the point of collision as seen in Fig. \ref{fig5}
where the distribution of the integrated population difference
$\widetilde{W}$ after the collision is plotted. For larger values of
$\Delta_0$, the FP soliton appears at the exit and its intensity
grows with increasing $\Delta_0$ as is clearly seen in Fig.
\ref{fig4}(a). Simultaneously, the excitation of medium diminishes:
obviously, the strong inhomogeneous broadening cannot provide
interpulse interaction large enough to effectively trap radiation.
It is also should be noted that the BP $3 \pi$ pulse also loses part
of its energy due to collision, since the peak intensity drops from
about $4.5 \Omega_0^2$ (see Fig. \ref{fig1} or \ref{fig2}) to
approximately $3 \Omega_0^2$ (at $\Delta_0=0.1$) and $2.5
\Omega_0^2$ (at $\Delta_0=3$) and grows again for larger broadenings
(see Fig. \ref{fig4}(b)). Nevertheless, the BP pulse is always
present at the exit of the medium and can be considered as means for
controlling less intensive FP soliton.

\begin{figure}[t!]
\centering \includegraphics[scale=1, clip=]{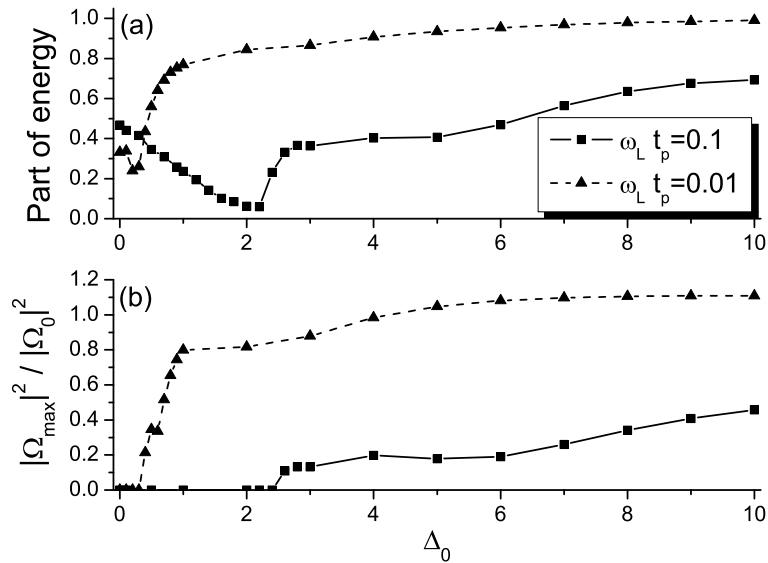}
\caption{\label{fig6} Dependence of (a) part of energy of $2 \pi$
pulse transmitted in forward direction (by the instant $100 t_p$)
and (b) the peak intensity of transmitted soliton on the level of
inhomogeneous broadening $\Delta_0$. The curves are shown for two
strengths of light-matter coupling.}
\end{figure}

These relationships can be traced in more detail in Fig. \ref{fig6}
where one can see the dependencies of the part of the FP pulse
energy transmitted through the medium and of the peak intensity of
the transmitted soliton on the parameter of inhomogeneous broadening
$\Delta_0$. It should be said that the transmitted energy is present
not only in the form of soliton but also as unstructured radiation
(dispersive waves). The difference can be clearly seen since the SIT
solitons always have characteristic shape described by hyperbolic
secant. The curves for the two variants of light-matter coupling are
shown in Fig. \ref{fig6}. First, let us discuss the case of strong
coupling ($\omega_L t_p=0.1$) corresponding to the results
represented in Figs. \ref{fig4} and \ref{fig5}. The soliton is
absent at the exit at low values of broadening ($\Delta_0 \lesssim
2.5$), though the substantial part of its energy (up to $50 \%$) is
transmitted in the form of low-intensity oscillations like those
discussed above. The minimal value of transmitted energy (only about
$6 \%$) is observed at $\Delta_0 \simeq 2.2$. Further increase in
broadening leads to the rapid jump of transmitted energy
corresponding to appearance of the single solitary pulse at the
exit. Subsequently monotonous growth of transmitted energy and peak
intensity of soliton occurs as is clearly demonstrated in Fig.
\ref{fig6}.

\begin{figure}[t!]
\centering \includegraphics[scale=1, clip=]{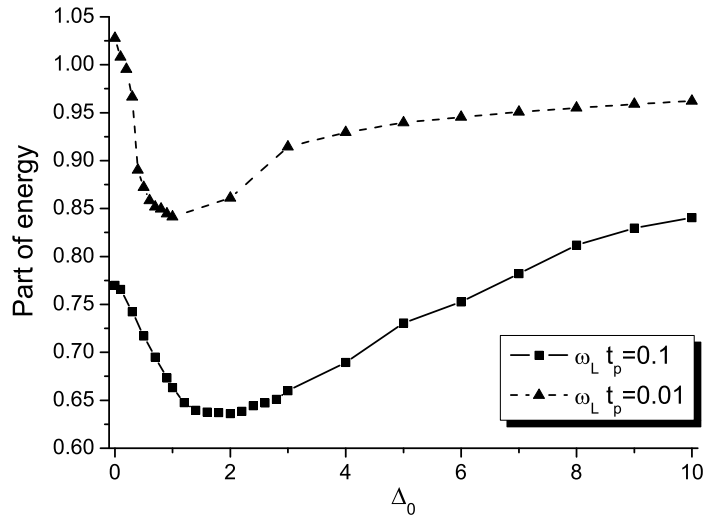}
\caption{\label{fig7} The same as in Fig. \ref{fig6}(a) but for $3
\pi$ pulse transmitted in backward direction.}
\end{figure}

Weakening the light-matter interaction ($\omega_L t_p=0.01$) results
in the pronounced shrinkage of the range of broadenings where the
transmitted soliton is absent. At the same time, the minimum of
transmitted energy shifts to the lower value of $\Delta_0$ and has
significantly larger magnitude. At large broadenings, the pulses
experienced the collision are more intensive than in the previous
case and have the parameters close to the nonperturbed solitons.
Finally, Fig. \ref{fig7} demonstrates similar dependencies of the
transmitted energy for the BP $3 \pi$ pulse. It is seen that
increase in the light-matter coupling results in reinforcement of
inelasticity, since the pulse loses considerably more energy due to
collision at $\omega_L t_p=0.1$ than in the case of $\omega_L
t_p=0.01$. It should be also noted that the position of the minimum
at $\omega_L t_p=0.1$ approximately coincides with that of Fig.
\ref{fig6}(a); it is not so evident for $\omega_L t_p=0.01$.
Nevertheless, one can say that there exists the optimal
inhomogeneous broadening for observation of inelastic soliton
collisions.

\section{\label{concl}Conclusion}

This study of coherent pulse propagation in the two-level medium is
focused on the influence of inhomogeneous broadening on the SIT
soliton formation and on the collisions of solitons. Though the main
features of these processes remain unchanged, there are some
interesting details which distinguish the case of strongly
inhomogeneous broadening from its homogeneous counterpart. Formation
of the SIT soliton in the homogeneously broadened medium is
accompanied by the characteristic oscillations of pulse intensity
(and its area), so that there is a distance of optimal compression
of the pulse. At this distance, the pulse is more intensive than
after ending of the formation process. In strongly inhomogeneously
broadened case, these oscillations are entirely washed out and,
hence, the pulse cannot be compressed more than it occurs in the
final state of SIT soliton. The inelastic collisions of
counter-propagating SIT solitons has the optimal conditions as well:
there is a nonzero level of inhomogeneous broadening which provides
the optimal trapping of radiation by the medium due to the
collision. Further increasing the width of broadened line or
decreasing the light-matter coupling strength (e.g. due to diluting
the medium) results in sharp rising of elasticity of collisions. We
believe that the results of this study will be helpful for
understanding of some subtle details of light-matter interaction.

\end{document}